\documentclass[aip,amsmath,amssymb,reprint,]{revtex4-1}
\usepackage[pdftex]{graphicx}

\begin{document}

\title{Power-law decay in the nonadiabatic photodissociation dynamics of alkali halides due to quantum wavepacket interference}

\author{Yuta Mizuno}
\email{mizuno@huku.c.u-tokyo.ac.jp}
\affiliation{
Department of Basic Science, Graduate School of Arts and Sciences, The University of Tokyo, 3-8-1 Komaba, Meguro-ku, Tokyo 153-8902, Japan
}

\author{Koji Hukushima}
\affiliation{ 
Department of Basic Science, Graduate School of Arts and Sciences, The University of Tokyo, 3-8-1 Komaba, Meguro-ku, Tokyo 153-8902, Japan
}
\affiliation{Research and Services Division of Materials Data and Integrated System, National
Institute for Materials Science, 1-2-1 Sengen, Tsukuba, Ibaraki 305-0047, Japan}

\date{\today}

\begin{abstract}
The nonadiabatic photodissociation dynamics of alkali halide molecules excited by a femtosecond laser pulse in the gas phase are investigated theoretically,
and it is shown that the population of the photoexcited molecules exhibits power-law decay with exponent $-1/2$,
in contrast to exponential decay, which is often assumed in femtosecond spectroscopy and unimolecular reaction theory.
To elucidate the mechanism of the power-law decay, a diagrammatic method that visualizes the structure of the nonadiabatic reaction dynamics as a pattern of occurrence of dynamical events, such as wavepacket bifurcation, turning, and dissociation, is developed.   
Using this diagrammatic method, an analytical formula for the power-law decay is derived,
and the theoretical decay curve is compared with the corresponding numerical decay curve computed by a wavepacket dynamics simulation in the case of lithium fluoride.
This study reveals that the cause of the power-law decay is the quantum interference arising from the wavepacket bifurcation and merging due to nonadiabatic transitions.
\end{abstract}

\maketitle

\begin{figure}[tb]
\begin{center}
\includegraphics[scale=1.0,clip]{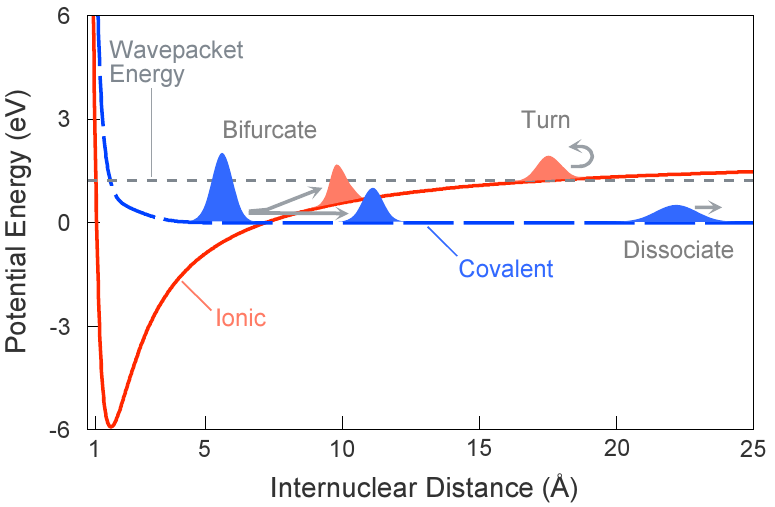}
\end{center}
\caption{Schematic illustration of the photodissociation dynamics of alkali halides, in the case of LiF as a typical example. 
The excited wavepackets repeat bifurcation, turning, and dissociation on the ionic and covalent diabatic potential energy curves.}
\label{Fig.1}
\end{figure}

The photodissociation dynamics of alkali halides, such as NaI and LiF, are a textbook example of nonadiabatic reaction dynamics.\cite{Rose1988,Rose1989,Cong1990,Knopp1997,Schmidt2015,Rasskazov2017,Kono1991,Dietz1996,Cornett1999,Balakrishnan1999,Alekseyev2000,Moeller2000,Takatsuka2015} 
The photodissociation dynamics are described as nuclear wavepacket dynamics on diabatic potential energy curves (PECs) of ionic and covalent electronic states, as shown in Fig.~\ref{Fig.1}.
After a ground-state wavepacket is pumped into the covalent state by a femtosecond laser pulse, the photoexcited wavepacket travels toward dissociation and reaches the crossing point of the PECs, where the wavepacket bifurcates into two components on the ionic and covalent PECs due to nonadiabatic transition.
The covalent component continues to travel toward dissociation, while the ionic component turns at the outer classical turning point. 
After turning, the ionic component reaches the crossing point and bifurcates into two components again, 
and the two bifurcated components turn at the inner classical turning points on each PEC.
In this way, the wavepackets undergo bifurcation, turning, and dissociation repeatedly, and the population of the photoexcited molecules that have not dissociated decreases with time. 

The population decay of the photoexcited alkali halide molecules is often assumed to be single-exponential or bi-exponential in previous experimental studies of femtosecond spectroscopy.\cite{Rose1989,Cong1990,Knopp1997,Schmidt2015,Rasskazov2017}
Exponential decay is also assumed in classic theory of unimolecular reaction, where unimolecular reaction kinetics are often modeled by a linear differential equation of the reactant population.

In contrast to these phenomenological assumptions, modern theory of unimolecular reaction reveals that the population decay does not always exhibit such simple exponential laws.
\cite{Moeller2000,Fonda1978,Shapiro1972,Shapiro1998,Nordholm1975,Estrada1989,Gertitschke1993,Desouter1997,Brems2002}
Time-independent theory of scattering shows that the spectral line shapes are deviate from the Lorentzian form due to strong coupling between bound (ionic) and continuum (covalent) states, and due to interference between overlapping resonances.\cite{Shapiro1972} 
Time-dependent theory of scattering shows that the dynamics of bound states are described formally by coupled integro-differential equations, which include memory effects.\cite{Nordholm1975,Estrada1989,Gertitschke1993,Desouter1997,Brems2002}  
These general facts imply that the population decay can exhibit non-exponential behavior.

In particular, Balakrishnan {\it et al.}\ showed, by numerical simulation, that the time-dependent population of excited LiF molecules, $P(t)$, exhibits a power-law with exponent $-1/2$ in time $t$:
$P(t) \propto t^{-\frac{1}{2}}$.\cite{Balakrishnan1999} 
Balakrishnan {\it et al.}\ explained qualitatively that the power-law decay is due to a coherent superposition of long-lived resonance states formed by interference between traveling waves in the ionic and covalent channel.\cite{Balakrishnan1999}
The power-law behavior can be interpreted qualitatively also in terms of scattering theory mentioned above. 
However, the detailed mechanism and emergence condition of the power-law decay have not been clarified quantitatively yet. 

In this communication, we elucidate the mechanism of the power-law decay by a diagrammatic approach based on the perspective of coherent wavepacket motion, different from the conventional scattering theoretical approach.
We derive an analytical formula for the population decay, which indicates the power-law with the characteristic exponent $-1/2$ explicitly,  
and clarify the condition for the emergence of the power-law decay. 
We also give a numerical demonstration, using the LiF model shown in Fig.~\ref{Fig.1}.

\begin{figure*}[tb]
\begin{center}
\includegraphics[scale=1.0,clip]{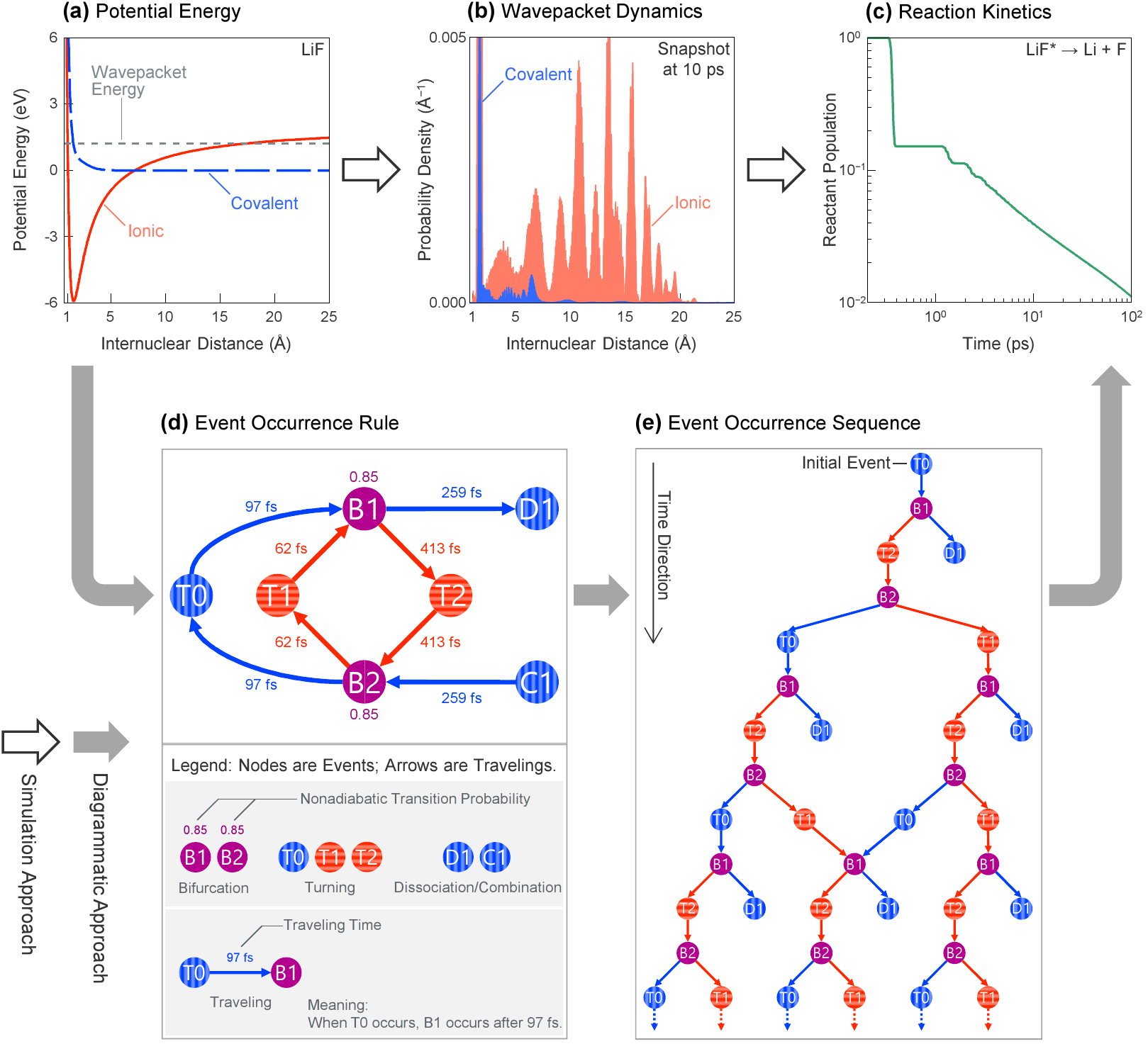}
\end{center}
\caption{Strategy for elucidating the mechanism of the power-law decay: a conventional simulation approach and an alternative diagrammatic approach.
Panel (a) shows the potential energy curves of the LiF model. 
Panels (b) and (c) show a snapshot of the wavepackets (probability density) and a decay curve of the reactant population, computed by a wavepacket dynamics simulation, respectively.
The diagram in panel (d) represents the rule of occurrence of dynamical events, determined by the potential energy curves and the wavepacket energy shown in panel (a).
This diagram includes all possible events at this wavepacket energy:
the bifurcations in the outward direction (B1) and in the inward direction (B2);
the turnings at the covalent inner turning point (T0), at the ionic inner turning point (T1), and at the ionic outer turning point (T2);
the dissociation (D1) and the combination (C1).
The diagram in panel (e) represents the sequence of occurrence of dynamical events.
This is obtained by tracing arrows in panel (d) from the initial event and merging paths that reach the same event at the same time.
In the case of the photodissociation process, the initial event is T0, and the combination event C1 never occurs.
By contrast, the initial event is C1 in the case of the atomic collision process $\mathrm{Li}+\mathrm{F}\to\mathrm{LiF}^\ast$.
A merging of paths in the sequence diagram represents an inevitable merging of wavepackets, which causes quantum interference.
}
\label{Fig.2}
\end{figure*}

The strategy we take is illustrated by Fig.~\ref{Fig.2}.
Conventional theoretical or numerical approaches to investigate the photodissociation dynamics include scattering theory and wavepacket dynamics simulation.  
The conventional scattering theorical approach is based on energy spectra\cite{Shapiro1972,Shapiro1998} or on coupled integro-differential equations\cite{Nordholm1975,Estrada1989,Gertitschke1993,Desouter1997,Brems2002}.
In the case of the photodissociation dynamics of alkali halides, the energy spectra have a complex, distorted line shape \cite{Cornett1999,Shapiro1998} and time-domain analysis based on them is cumbersome.
The coupled integro-differential equations have long-time memory terms and difficult to solve theoretically.\cite{Estrada1989,Gertitschke1993,Desouter1997,Brems2002} 
Wavepacket dynamics simulations provide snapshots (or movies) of wavepacket dynamics, as shown in Fig.~\ref{Fig.2}(b). 
These graphics are also complex and difficult to interpret, because the wavepackets repeatedly bifurcate, merge, and interfere with each other.\cite{Balakrishnan1999,Takatsuka2015}
Thus we developed an alternative method based on the diagrams shown in Figs.~\ref{Fig.2}(d) and \ref{Fig.2}(e).
These diagrams visualize the structure of the complicated dynamics as a pattern of occurrence of dynamical events, such as wavepacket bifurcation, turning, and dissociation.

The diagram in Fig.~\ref{Fig.2}(d) represents the rule of event occurrences. 
This diagram is determined by the PECs and the wavepacket energy shown in Fig.~\ref{Fig.2}(a).
In this diagram, all possible events at this wavepacket energy are depicted by nodes,
and wavepacket travelings between these events are depicted by arrows.
This diagram also includes quantitative dynamical parameters as follows: 
(i) Nonadiabatic transition probabilities, which can be calculated using the Landau--Zener formula etc., are assigned to each bifurcation node; 
(ii) Traveling times, which can be calculated approximately as the times it takes a classical particle to travel between events, are assigned to each arrow.\cite{Footnote_NotMarkov}
In addition to the above parameters, the quantum mechanical phase must be considered in rigorous quantum or semiclassical treatment.
However, the phase parameters do not affect the power-law decay quantitatively, as described below.

The diagram in Fig.~\ref{Fig.2}(e) represents the sequence of event occurrences.
This diagram is obtained by tracing arrows in the event-occurrence-rule diagram from the initial event and merging paths that reach the same event at the same time.
In the case of the photodissociation dynamics, the initial event is T0, because the photoexcited wavepacket is at the covalent inner turning point immediately after excitation.
A merging of paths in this diagram represents an {\it inevitable} merging of wavepackets, which causes quantum interference. 
For example, there is a merging point B1 at the central lower part of Fig.~\ref{Fig.2}(e). 
The sets of events on the merged paths from the initial event to the merging event are the same as $\{\mathrm{T0\times2, T1\times1, T2\times2, B1\times3, B2\times2}\}$,
while the occurrence orders of the events are different.
This results in the same traveling time of the two different paths and the inevitable merging of the wavepackets that travel along the two paths. 

The structure of bifurcation and merging depicted in Fig.~\ref{Fig.2}(e) leads to the power-law decay with exponent $-1/2$. We explain the detailed mechanism in four steps.

\begin{figure}[ptb]
\begin{center}
\includegraphics[scale=1.0,clip]{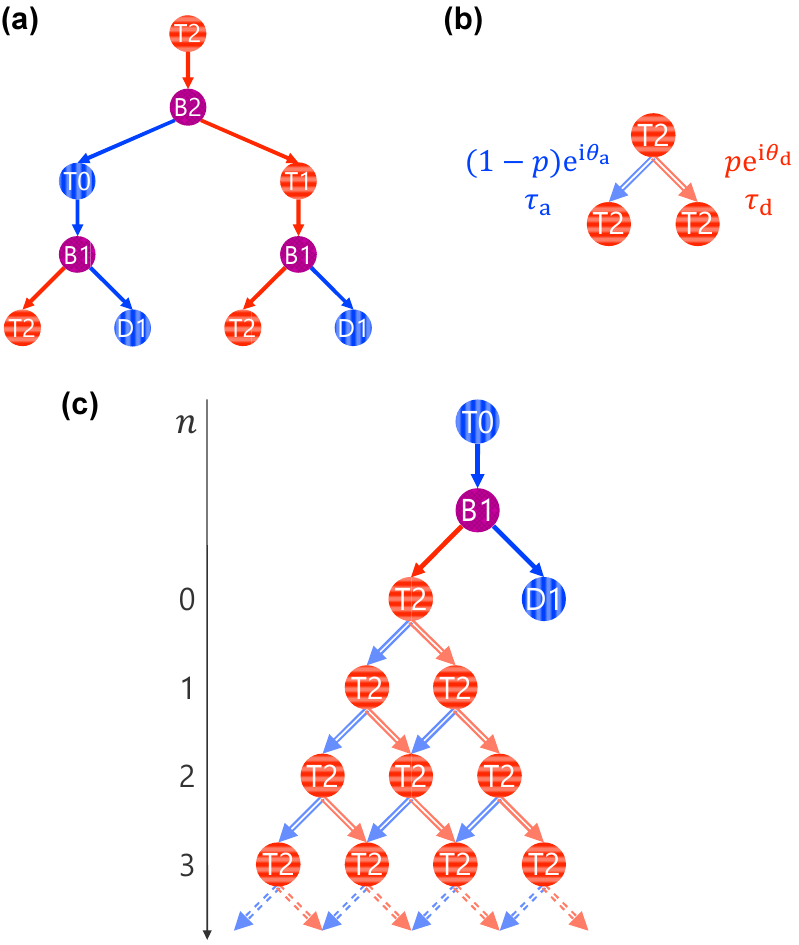}
\end{center}
\caption{Structure of Pascal's triangle hidden in the sequence of event occurrences, which leads to the power-law decay with exponent $-1/2$. Expressing a building-block diagram (a) by a simplified diagram (b), we can diagram the sequence of event occurrences as the diagram (c). In panel (b), $p$ is the nonadiabatic transition probability, $\theta_{\mathrm{a}}$ and $\theta_{\mathrm{d}}$ are the phases acquired along each path, and
$\tau_{\mathrm{a}}$ and $\tau_{\mathrm{d}}$ are the traveling times of each path.
In panel (c), $n$ denotes the row number of the Pascal's triangle.   
}
\label{Fig.3}
\end{figure}

The first step is to determine the hidden structure in Fig.~\ref{Fig.2}(e) that leads to the power-law decay. 
To clarify this structure, we introduce a building-block diagram. 
The whole sequence has a recurrence structure of a building block shown in Fig.~\ref{Fig.3}(a).
This building block corresponds to one cycle of nuclear vibration.
We express the building block by a simplified diagram shown in Fig.~\ref{Fig.3}(b).
Here, $p$ is the nonadiabatic transition probability, $\tau_{\mathrm{a}}$ is the traveling time of the adiabatic path (the path on the excited adiabatic PEC via the event T0), $\theta_{\mathrm{a}}$ is the phase acquired along the adiabatic path, $\tau_{\mathrm{d}}$ is the traveling time of the diabatic path (the path on the ionic diabatic PEC via the event T1), and $\theta_{\mathrm{d}}$ is the phase acquired along the diabatic path.
Note that the complex parameters $(1-p)\mathrm{e}^{\mathrm{i}\theta_{\mathrm{a}}}$ and $p\mathrm{e}^{\mathrm{i}\theta_{\mathrm{d}}}$ are the probability amplitudes of travelings along the adiabatic and diabatic path, respectively.
Using this simplified building-block diagram, the sequence of event occurrences is diagrammed as shown in Fig.~\ref{Fig.3}(c).
Because the first part of the sequence is impossible to express by the building-block diagram, the top of this diagram is drawn in the original manner.
The structure of this diagram, except the top part, is Pascal's triangle and leads to the power-law decay. 

The second step is to calculate the occurrence probability amplitude of each event in the Pascal's triangle.
The Pascal's triangle structure yields a binomial distribution of the occurrence probability amplitudes.
The occurrence amplitude of the event of the $n$-th row and $k$-th column of the Pascal's triangle is 
\begin{equation}
 \psi^n_k = \binom{n}{k} p^k(1-p)^{n-k} \mathrm{e}^{\mathrm{i}\theta^n_k}, \label{Eq.binomial}
\end{equation}
where $\binom{n}{k}$ is the binomial coefficient, 
\begin{equation}
 \theta^n_k = k\theta_{\mathrm{d}} + (n-k)\theta_{\mathrm{a}}, 
\end{equation}
and $\psi_0^0$ is set to 1.
Here, the quantum interference arising from the merging of paths is taken into account by the coherent superposition
\begin{equation}
\psi^n_k = p\mathrm{e}^{\mathrm{i}\theta_{\mathrm{d}}}\psi^{n-1}_{k-1} + (1-p)\mathrm{e}^{\mathrm{i}\theta_{\mathrm{a}}}\psi^{n-1}_k.\label{Eq.CohSup}
\end{equation}
Since the binomial distribution is approximated by the normal distribution for large $n$,
the occurrence amplitude is approximated by
\begin{equation}
\psi^n_k \approx \frac{1}{\sqrt{2\mathrm{\pi}\sigma_n^2}}\exp\left[-\frac{(k-np)^2}{2\sigma_n^2}+\mathrm{i}\theta^n_k\right],
\end{equation}
where
\begin{equation}
 \sigma_n = \sqrt{np(1-p)}.    
\end{equation}

The third step is to calculate the population of undissociated molecules on the $n$-th cycle of nuclear vibration.
Assuming that different events in the Pascal's triangle do not occur simultaneously,
the population on the $n$-th cycle can be approximated by the sum of the occurrence probabilities of the events in the $n$-th row of the Pascal's triangle:
\begin{eqnarray}
P_n &\approx& \sum_{k=0}^n |\psi_k^n|^2 \label{Eq.P_n_1}\\
&\approx& \frac{1}{2\mathrm{\pi}\sigma_n^2}\sum_{k=0}^n \exp\left[-\frac{(k-np)^2}{\sigma_n^2}\right]. \label{Eq.P_n}
\end{eqnarray}
Approximating the sum in Eq.~(\ref{Eq.P_n}) by the Gaussian integral, we obtain
\begin{equation}
 P_n \approx \frac{1}{2\mathrm{\pi}\sigma_n^2}\int_{-\infty}^{\infty}\exp\left[-\frac{x^2}{\sigma_n^2}\right]\mathrm{d}x
 =
 \frac{1}{\sqrt{4\mathrm{\pi}\sigma_n^2}}.\label{Eq.P_n_Gauss}
\end{equation}
Note that the phases $\theta_{\mathrm{a}}$ and $\theta_{\mathrm{d}}$ disappear in Eqs.~(\ref{Eq.P_n_1})--(\ref{Eq.P_n_Gauss}).
This is because of the above assumption that different events in the Pascal's triangle do not occur simultaneously;
in other words, due to the assumption that there is no quantum interference except that arising from the inevitable merging of wavepackets which travel along different paths with the same traveling time and phase.
If two different events in the Pascal's triangle have close occurrence times,
the two wavepackets that undergo each event can overlap each other due to their non-zero wavepacket widths.
Here we use the term ``overlapping'' as spatial overlapping of wavepackets, different from that used in scattering theory as mentioned in introduction. 
This {\it accidental} overlapping causes quantum interference, depending quantitatively on the acquired phases.
The accidental interference effect is described by the cross term, ${\psi_k^n}^\ast\psi_l^m$, which is neglected in Eq.~(\ref{Eq.P_n_1}).
The assumption of neglecting the accidental overlapping of wavepackets can be justified in the case that the pump pulse is short enough that the spatial width of the photoexcited wavepacket is sufficiently narrow. 

The final step is to calculate the population in the real-time domain.
The population $P_n$ is approximately equal to the population at around the time $n\bar{\tau}$, where 
\begin{equation}
\bar{\tau} = p\tau_{\mathrm{d}} + (1-p)\tau_{\mathrm{a}}
\end{equation}
is the mean traveling time of one cycle of vibration. 
Thus we can approximate the time-dependent population by
\begin{equation}
 P(t) \approx \frac{1}{\sqrt{4\mathrm{\pi}p(1-p)}}\sqrt{\frac{\bar{\tau}}{t}}.
\end{equation}
Considering the correction due to the top part of the diagram in Fig.~\ref{Fig.3}(c), we obtain the final expression:
\begin{equation}
 P(t) \approx \frac{P_0}{\sqrt{4\mathrm{\pi}p(1-p)}}\sqrt{\frac{\bar{\tau}}{t-t_0}},   \label{Eq.Final}
\end{equation}
where $P_0$ is the population of the molecules that undergo the event T2 at the top of the Pascal's triangle, and $t_0$ is the time when the first dissociation event D1 occurs. 
Equation~(\ref{Eq.Final}) indicates that the population decay exhibits a power-law with exponent $-1/2$.
This analytical formula is derived under two major assumptions:
(i) $t\sim n$ is large;
(ii) no accidental overlapping of wavepackets occurs.

\begin{figure}[ptb]
\begin{center}
\includegraphics[scale=1.0,clip]{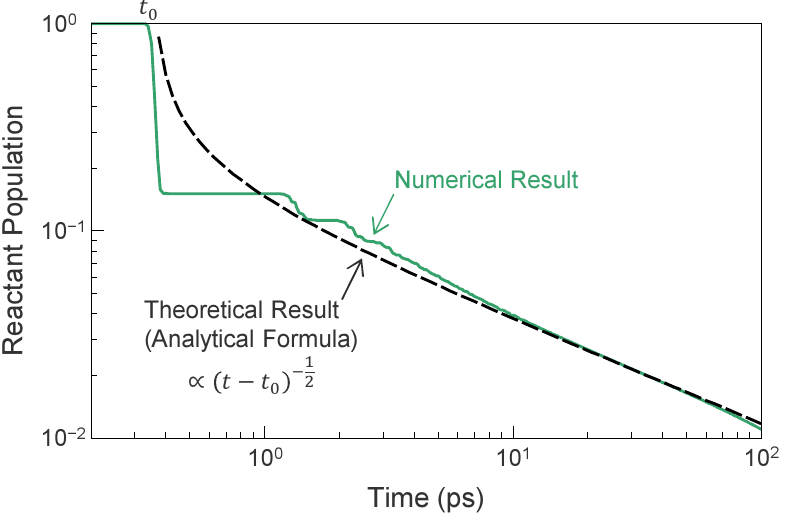}
\end{center}
\caption{Time-dependency of the reactant population in the LiF photodissociation.
The green-solid line is a numerical decay curve computed by a wavepacket dynamics simulation, and the black-dashed line is the corresponding theoretical decay curve of Eq.~(\ref{Eq.Final}), derived by the diagrammatic method. Here, the initial population is normalized to 1, and $t_0$ is the time when the first dissociation event occurs.
}
\label{Fig.4}
\end{figure}

To validate the above theory, we conduct a numerical demonstration. 
Figure~\ref{Fig.4} shows a numerical decay curve of the reactant population computed by a wavepacket dynamics simulation and the corresponding theoretical decay curve of Eq.~(\ref{Eq.Final}).
In this demonstration, the reactant population is defined as the population of photoexcited LiF molecules whose nuclear distances are less than $24.7~ \mathrm{\AA}$.
The initial state is the ground vibrational state on the ground electronic state, and it is pumped into the covalent state at 0 fs by a Gaussian pulse with center frequency 7.08 eV, peak intensity $10^{13}~ \mathrm{W/cm^2}$, and pulse width 20 fs.
In this case, the total energy of the excited wavepacket is 1.21 eV.  
The numerical curve was computed by the split-operator method\cite{Feit1982}, using {\it ab initio} PECs\cite{Giese2004} and an absorbing boundary condition\cite{Neuhasuer1989}.
The nonadiabatic transition probability $p$ was calculated using the Landau--Zener formula, and the traveling times were calculated by a classical dynamics simulation,
supposing that the energy of the nuclear motion is 1.21 eV.
These parameters are shown in Fig.~\ref{Fig.2}(d).
Note that the width of the pump pulse (20 fs) is sufficiently shorter than the difference between the traveling times of the adiabatic and diabatic paths (70 fs).
Therefore the accidental overlapping of the wavepackets that undergo different events in the same row of the Pascal's triangle can be neglected.

In Fig.~\ref{Fig.4}, the theoretical decay curve is in good agreement with the numerical decay curve in the long-time region ($>$ 10 ps).
Since there is no adjustable parameter in the analytical formula of Eq.~(\ref{Eq.Final}), this agreement between the theoretical and numerical results is strong evidence that the population of the photoexcited LiF exhibits power-law decay with exponent $-1/2$.

The power-law decay should emerge also in the case of other alkali halides.
Alkali halides with a light halogen atom, such as LiCl, have PECs with the same characteristic as LiF; that is, one covalent-dissociative PEC interacts with one ionic-binding PEC.\cite{Giese2004,Weck2004} 
In those cases, the above analysis is valid, and the population decay should exhibit power-law with exponent $-1/2$.
By contrast, alkali halides with a heavy halogen atom, such as NaI, have PECs with a different characteristic from that of LiF;
{\it two} covalent-dissociative PECs, corresponding to  two spin-orbit states of the heavy halogen atom, interact with one ionic-binding PEC.\cite{Alekseyev2000}
In those cases, the above analysis is valid only when the wavepacket energy is low enough that the upper covalent PEC is not accessible energetically.  
When both covalent PECs are accessible, the above analysis, based on the Pascal's triangle and the binomial distribution, cannot be adopted,
and the population decay may not exhibit the same power-law as for LiF.   
However, the scheme of the diagrammatic approach shown in Fig.~\ref{Fig.2} is still valid, and an analysis using a multinomial distribution may be feasible.

The quantum interference arising from the inevitable merging of wavepackets is necessary for the power-law decay.
If there is no quantum interference, that is, the merging of paths is described by the incoherent superposition
\begin{equation}
\left|\psi^n_k\right|^2 = p^2\left|\psi^{n-1}_{k-1}\right|^2 + (1-p)^2\left|\psi^{n-1}_k\right|^2 \label{Eq.IncohSup}
\end{equation}
instead of the coherent superposition of Eq.~(\ref{Eq.CohSup}),
the population decay is exponential.
This is because Equation~(\ref{Eq.IncohSup}) is linear with respect to occurrence probabilities.
By contrast, the coherent decay process cannot be described by a linear equation due to the interference term
${\psi_{k-1}^{n-1}}^\ast \psi_k^{n-1}$.
Thus the quantum interference causes non-exponential behavior, i.e. power-law decay in the present case.
Here, note that the quantum interference does not arise before two cycles of vibration.
Therefore the population decay for the short time can be described as single-exponential decay.
This is consistent with previous studies\cite{Balakrishnan1999,Moeller2000}.  
 
In the field of quantum dynamics, many studies\cite{Kono1991,Dietz1996,Granucci1995,Romstad1997,Shapiro1999,Zhang2003,Gador2004} pointed out that wavepacket interference plays an important role in nonadiabatic reaction dynamics.
These studies, however, focused on only accidental interference depending on quantitative parameters, such as wavepacket energy.
In contrast to this, the present study focuses on the inevitable interference not depending on quantitative parameters.
In the field of nonlinear dynamics, Takatsuka and coworkers showed that the bifurcation and merging of wavepackets due to nonadiabatic transitions induce ``quantum chaos," from the perspective of energy-level statistics\cite{Fujisaki2001} and state relaxation\cite{Higuchi2002}.
This ``quantum chaos" and the power-law behavior we showed are nonlinear aspects of nonadiabatic dynamics.
This study revealed a new aspect of quantum and nonlinear effects in nonadiabatic reaction dynamics from the point of view of coherent wavepacket motion and its structure. 

The diagrammatic method we developed provides an intuitive interpretation of the complicated nonadiabatic reaction dynamics of alkali halides
and enables us to elucidate the detailed mechanism of the power-law decay quantitatively.
However, the present theory neglects the accidental overlapping of wavepackets and its quantum interference effect.
This accidental overlapping is likely to occur in extremely long-time regions owing to the following mechanism:
the deviation of the occurrence times of the events in the same row of the Pascal's triangle increase with time,
and occurrence times of events in adjacent rows can be accidentally close to each other in extremely long-time regions;
this leads to the accidental overlapping of wavepackets of adjacent cycles of nuclear vibration, even though the wavepackets of the same cycle do not overlap.
Indeed, the theoretical decay curve in Fig.~\ref{Fig.4} is in slight disagreement with the numerical counterpart in the extremely long-time region ($>$ 80 ps). 
Therefore it is necessary to extend the present theory to include the influence of the accidental overlapping.

In summary, using the diagrammatic method, we showed that the population of photoexcited alkali halides exhibits power-law decay with exponent $-1/2$ due to the quantum interference arising from the inevitable merging of wavepackets.
The condition for the emergence of the power-law decay is that accidental overlapping of wavepackets can be neglected.
This power-law behavior was first pointed out by Balakrishnan {\it et al.}\cite{Balakrishnan1999},
and we revealed its detailed mechanism and emergence condition by the diagram-based analysis. 
We will present the detailed formulation of the diagrammatic method, including a semiclassical description of the quantum phases, in a separated paper.
We will also report the further investigation into the population decay of photoexcited alkali halides, including the case of alkali halides with a heavy halogen atom and the influence of the accidental overlapping, in the future.

We are grateful to Dr. Yasuki Arasaki for providing the data on the potential energy curves interpolated on the basis of the {\it ab initio} data in Ref.~\onlinecite{Giese2004}.    
This work was supported by JSPS KAKENHI Grant Number 17J10744 (to Y.M.).
Y.M. is a research fellow of the Japan Society for the Promotion of Science.


\begin{thebibliography}{99}
\bibitem{Footnote_NotMarkov}
Note that the traveling time is not a kind of lifetime nor a kind of inverse of a rate constant, but the time interval between event occurrences.
In other words, this diagram does not represent a Markov process model nor a network of reaction kinetics, but an abstract model of reaction dynamics.
\end{thebibliography}
\end{document}